\documentclass[11pt]{article}
\usepackage{geometry}                
\usepackage[parfill]{parskip}    
\usepackage{graphicx}
\usepackage{amsmath, amsthm, amsfonts}
\usepackage{amssymb}

\usepackage{color}
     \usepackage[colorlinks]{hyperref}
\hypersetup{linkcolor=blue,%
citecolor=red,%
urlcolor=cyan}
\usepackage{url} 

\usepackage{epstopdf}
\begin{document}

\title{Graphene Dirac fermions in symmetric
electric and  magnetic fields: 
\\
the case of an electric square well}
 
 \author{\.Ismail Burak Ate\c{s}$^a$\footnote{ismailburakates@gmail.com, ORCID: \href{http://orcid.org/0000-0001-8262-2572}{0000-0001-8262-2572}}, 
\c Seng\"ul Kuru$^a$\footnote{sengul.kuru@science.ankara.edu.tr, ORCID: \href{http://orcid.org/0000-0001-6380-280X}{0000-0001-6380-280X}}, 
Javier Negro$^b$\footnote{jnegro@uva.es, 
ORCID: \href{http://orcid.org/0000-0002-0847-6420}{0000-0002-0847-6420}}
\\ 
$^a$Department of Physics, Faculty of Science, Ankara University, 06100 Ankara, Turkey\\
$^b$Departamento de F\'i{}sica Te\'orica, At\'omica y Optica, Universidad de Valladolid,\\ 47071 
Valladolid, Spain\\}

	\maketitle
	
\begin{abstract}
In this paper, a simple method is proposed to get analytical solutions (or with the help of a finite numerical calculations) of the Dirac-Weyl equation for low energy  electrons in graphene in the presence of certain  electric and magnetic fields. 
In order to decouple the Dirac-Weyl equation  we have assumed a displacement symmetry of the system along a direction and some conditions on the magnetic and electric fields. The resulting equations have the natural form to apply the technique of supersymmetric quantum mechanics. The example of an electric well with square profile is worked out in detail to illustrate some  of the most interesting features of this procedure.
\end{abstract}

\section{Introduction}
Graphene being the first example of two-dimensional crystals  has attracted much attention in physics due to its important electronic and magnetic 
properties \cite{Katsnelson, Neto,  Goerbig}. 
The dynamics of low-energy, massless spin 1/2 Dirac electrons with Fermi-velocity in graphene is described by the (2+1) dimensional Dirac-Weyl equation \cite{Neto}. The analytical solutions of this equation in the presence of different external magnetic and electric fields and the interpretation of its results, has brought great interest in recent years for its applications to graphene and other allotropes of carbon. For instance, it is essential for a better knowledge of the electronic properties of nanoribbons, nanowires and nanotubes. In particular, studies on graphene quantum dots (also known as artificial atoms), 
contribute to basic physics, electronics, optics, etc. which have numerous applications \cite{portnoi3, Afshari, Do21}. Therefore, not only theoretical but also experimental studies are conducted to understand the confinement mechanism of Dirac electrons in graphene \cite{Freitag,Jiang}. Due to Klein tunnelling, it is known the difficulty to trap them by means of electrostatic potential \cite{Katsnelson,Hewageegana, Beenakker}; it is much easier in this respect to appeal to  magnetic fields either in graphene  \cite{Milpas,kuru, Negro18, Moldovan,
Van} or in other shapes such as spherical fullerenes \cite{Negro13, roy19}  or in hyperbolic   geometry \cite{ Negro20}. However, since electrostatic potential may be more advantageous in practice, many studies on the effect of electric fields have also been carried out recently \cite{Bardarson,Wang,Negro22,jakubsky22,superKlein,portnoi4,portnoi5,lukose,roy20}. 

The aim of this study is to search for analytical methods to obtain solutions, or approximations, to the trapping mechanisms of Dirac electrons in graphene under  magnetic and electric external fields. In the present work we will restrict to systems with displacement symmetry along one direction (the $y$ axis) to facilitate the approach.
The different conditions that we will find to classify these systems can be applied to many results scattered in the recent literature. Thus, we wish  they to be considered under the same point of view which consist in three options:  pure magnetic or electric field, or both mixed perpendicular electric and magnetic fields. We will deal in detail with one example of pure electric field in order to show some not so well known peculiarities: the role of the symmetry momentum $k_y$ in this problem; the electric potential as an effective complex potential or the  supersymmetric character of such complex potentials.

Supersymmetric quantum mechanics (SUSY-QM) has been shown to be a very fruitful tool for finding and analyzing exact solutions to some 1-dimensional problems \cite{Cooper,david10}. This method can also be applied to the 2D Dirac-Weyl equation, even for 2D surfaces with curvature \cite{kuru,Negro18,Negro13,Negro20,Fernandez,david22,schulze,Le19,Lee19}. Within the scope of this work, it is aimed to find and solve the problems involving electric and/or magnetic fields by means of this method. This requires a specific form of electric and magnetic fields; therefore, it might be difficult that these fields be directly applied in experimental studies. However, such problems are theoretically important and  can contribute to a more comfortable understanding of some of the facts encountered in practice. 

The organization of the paper is as follows. In Section 2 we present the general setup of 
the Dirac-Weyl equation interacting with electric and magnetic fields. A matrix $M$ is
introduced which determines three possible cases.  These cases are analysed in Section 3; they correspond to pure magnetic or electric fields and both fields together. The pure
electric field is carefully examined through an example in Section 4. The last section enumerates
the most interesting results of the work.

\section{Interactions with symmetry in the y-axis direction}
	
We will restrict here to the stationary Dirac-Weyl equation in the plane for quasi-particles of zero mass and $1/2$-spin but with the charge of an electron. This is the continuous approximation to describe the electronic interaction in graphene for a range of energies near  the Dirac points (hereafter we will restrict to one of them,  $K$). The equation for the interaction in external static fields, applying the minimal coupling rule, is
\begin{equation}\label{dirac1}
v_F\big(  {\boldsymbol \sigma \cdot}  ({\bf p}+e{\bf A}) \big)\Psi({\bf x}) = \big(E - V({\bf x})\big) \Psi({\bf x})
\end{equation}
where, as usual ${\boldsymbol \sigma} = (\sigma_x,\sigma_y)$ are sigma Pauli matrices,
$\bf p$ the two-component momentum operator in the $xy$-plane, $v_F$ the Fermi velocity of graphene, $e$ the electric charge and ${\bf A}({\bf x})$, $V({\bf x})$  magnetic and electric potentials defined on the plane; the wavefunction
$\Psi({\bf x})=\Psi(x,y)$ is a two-component pseudospinor \cite{Neto}. 

We will assume the symmetry of the fields along displacements in one direction (in the $y$-direction) and both fields to be perpendicular (the electric on the $xy$-plane, the magnetic in the $z$-direction perpendicular to the graphene plane). In fact, we will restrict to the potentials having the form
\begin{equation}
{\bf A}({\bf x}) = (0, A_y(x)),\qquad V({\bf x})= V(x)
\end{equation}
Therefore, the magnetic field may depend only on $x$ in this case and it is given by ${\bf B}(x)= A'_y(x) \hat z=\frac{dA_y(x)}{dx}\hat z$, where $\hat z$ is the unit vector of $z$-axis;
while the electric field ${\bf E}(x)= -V'(x) \hat x=-\frac{dV(x)}{dx}\hat x$ will have the $x$-direction and it may depend also on $x$. Hereafter, 
the ``prime", will be used for derivatives with respect to the argument.
Then, we can choose the wavefunction of (\ref{dirac1}) to be an eigenfunction of the translation operator $p_y = -i\hbar\partial_y$, with eigenvalue $\hbar k$,
\begin{equation}\label{psiy}
\Psi(x,y)= e^{i k \,y}\left(\begin{array}{c}
\psi_1(x)\\
i\psi_2(x) \end{array}\right),\qquad p_y \Psi(x,y) = \hbar k \Psi(x,y)
\end{equation}
where the imaginary unit $i$ has been included in the second component by convenience.
Replacing these fields and operators together in the initial equation (\ref{dirac1}), we have  a  matrix differential eigenvalue problem 
\begin{equation}\label{dirac2}
\hbar v_F
\left(\begin{array}{cc}
0 & \partial_x +k + e/\hbar\, A_y(x)\\[1.ex]
-\partial_x + k + e/\hbar A_y(x)& 0 \end{array}\right)
\left(\begin{array}{c}
\psi_1(x)\\[1.ex]
\psi_2(x) \end{array}\right)=
(E-V(x)) \left(\begin{array}{c}
\psi_1(x)\\[1.ex]
\psi_2(x) \end{array}\right)
\end{equation}

Next, we will manipulate this equation to see if it can be diagonalized (or decoupled). Thus, we rewrite the equation in the following form

\begin{equation}\label{dirac3}
\left(\begin{array}{cc}
\partial_x   &  0\\[1.ex]
 0 & \partial_x   \end{array}\right)
\left(\begin{array}{c}
\psi_1(x) 
\\[1.ex]
\psi_2(x) \end{array}\right) 
= \left(\begin{array}{cc}
( k + \frac{e}{\hbar}\, A_y(x)) & -(E-V(x))/\hbar v_F
\\[1.ex]
(E-V(x))/\hbar v_F & -(k + \frac{e}{\hbar}\, A_y(x))\end{array} \right)
\left(\begin{array}{c}
\psi_1(x)\\[1.ex]
\psi_2(x) \end{array}\right)
\end{equation}

We can simplify the notation as follows:
\begin{equation}\label{m1}
W(x) := k +e/\hbar\, A_y(x):=k + {\cal A}_y(x),
\qquad 
\Delta(x) := 
\frac{E}{\hbar v_F}-\frac{V(x)}{\hbar v_F}:=\varepsilon- {\cal V}(x), 
\end{equation}
so that $W(x)$ is related to the magnetic potential and $\Delta(x)$ with the
electric one.
This is the Dirac-Weyl equation in the normal form of a system of two differential equations (in other words, with only the derivative of the components to the left):
\begin{equation}\label{px}
 \partial_x \Psi_R(x) = M \Psi_R(x)
\end{equation}
where the $M=M(x)$ matrix and the column vector $\Psi_R(x)$ are
\begin{equation}\label{m}
M= \left(\begin{array}{cc}
W(x) & -\Delta(x)
\\[1.ex]
\Delta(x) & -W(x)\end{array} \right)\,,\qquad 
\Psi_R(x)= \left(\begin{array}{c}
\psi_1(x) 
\\[1.ex]
\psi_2(x)\end{array} \right)
\end{equation}
Notice that this system of two linear equations is real and we are looking for real square integrable solutions
$\Psi_R(x) = \Psi_R(x)^*$, 
\[
\psi_1(x)=  \psi_1(x)^*,\qquad \psi_2(x)= \psi_2(x)^* 
\]
Differentiating  equation (\ref{px}) gives
\begin{equation}\label{pxx}
 \partial_{xx} \Psi_R(x) = (M' + M^2) \Psi_R(x)
\end{equation}
where $M^2$ is diagonal, while $M'=\frac{dM(x)}{dx}$ carries the nondiagonal part:
\begin{equation}
M'= \left(\begin{array}{cc}
W'(x) & -\Delta'(x)
\\[1.ex]
\Delta'(x) & -W'(x)\end{array} \right) ,\qquad 
M^2= \Big(W(x)^2- \Delta(x)^2\Big)\, I
\end{equation}
In the following we will discuss the simplest options to diagonalize $M'$ which will allow us to decouple eq.~(\ref{pxx}).

\section{Solvable cases}
The formulation of the problem is as follows. We must diagonalize $M'\to  T^{-1}M'T$ by means
of a constant matrix,  $T$, in order that it should commute with the differential operator $\partial_{xx}$ in
equation (\ref{pxx}). In this way we will arrive to two differential equations, one for each component.
There are three situations where this is possible, as it will be seen below.

\subsection{Pure magnetic field ${\cal V}=0$}

In this first case, $\Delta'(x) = {\cal V'}(x) = 0$, that is, we have a pure magnetic field and $M'$, $M^2$ take the form
\[
M'= \left(\begin{array}{cc}
W'(x) & 0
\\[1.ex]
0 & -W'(x)\end{array} \right),\qquad 
M^2= \Big(W(x)^2- \varepsilon^2\Big)\, I
\]

This is the simplest case, where we have just a magnetic field perpendicular to the plane $XY$, determined by the component ${\cal A}_y$. Besides, the matrix $M'(x)$ is already diagonal, and
replacing this in eq. (\ref{pxx}) we get the explicit form of  the SUSY-QM formal identification of these two  equations:
\begin{equation}\label{pxx2}
  \left(\begin{array}{cc}
-\partial_{xx}  + W'(x) + W(x)^2 & 0
\\[1.ex]
0 & -\partial_{xx} - W'(x)+ W(x)^2\end{array} \right)\Psi_R(x) = \varepsilon^2 \Psi_R(x)
\end{equation}
where $W(x)$ plays the role of superpotential.
This case was exhaustively studied in previous references, 
including also a generalization to surfaces with curvature 
\cite{Goerbig,Milpas,kuru,Cooper,Fernandez}. 

\subsection{Pure electric field ${\cal A}(x) = 0$}

In this case equation (\ref{pxx}) takes the form
\begin{equation}\label{pxx3}
 \partial_{xx} \Psi_R(x) = \Delta'(x)\left(\begin{array}{cc}
0 & - 1
\\[1.ex]
1 & 0\end{array} \right)\Psi_R(x) + (k^2- \Delta(x)^2)\left(\begin{array}{cc}
1 & 0
\\[1.ex]
0 & 1\end{array} \right) \Psi_R(x)
\end{equation}
where the non-diagonal matrix is
\[
M'= \Delta'(x) \left(\begin{array}{cc}
0 & -1
\\[1.ex]
1 & 0\end{array} \right)
=
{\cal V}'(x)
\left(\begin{array}{cc}
0 & 1
\\[1.ex]
-1 & 0\end{array} \right)
\]
 The eigenvalues of the  matrix  $M'$ are complex so that it can be transformed  in a diagonal form $\tilde M'$ by means of a complex matrix $T$:
\[
\tilde M'= T^{-1} M' T=
\left(\begin{array}{cc}
-i\,{\cal V}'(x) & 0
\\[1.ex]
0 & i\,{\cal V}'(x)\end{array} \right),
\qquad T=
\left(\begin{array}{cr}
1 & 1
\\[1.ex]
i & -i \end{array} \right),
\quad T^{-1}=
\frac{1}{2}\left(\begin{array}{cr}
1 & -i
\\[1.ex]
1 & i \end{array} \right)
\]
Then, defining 
\begin{equation}\label{p12}
\tilde \Psi = T^{-1} \Psi_R \ \implies 
\left\{\begin{array}{ll}\tilde \psi_1 = \frac12(\psi_1 -i \psi_2)
\\[1.ex]
\tilde \psi_2 = \frac12(\psi_1 +i \psi_2)
\end{array}\right.
\end{equation}
the system of equations (\ref{pxx3}) becomes 
\begin{equation}\label{pxxx2}
 \partial_{xx} \tilde\Psi(x) = (\tilde M' + \tilde M^2) \tilde\Psi(x)
\end{equation}
which consists in two complex conjugate equations, one for each component:
\begin{equation}\label{sch}
\left\{
\begin{array}{l}
\Big(-\partial_{xx}  + V_1(x)\Big) \tilde\psi_1 := \Big(-\partial_{xx}   -\big(i\,{\Delta}\big)'  + \big(i\,{\Delta}\big)^2\Big) \tilde\psi_1 = -  k^2\, \tilde\psi_1
\\[2.ex]
\Big(-\partial_{xx}  + V_2(x)\Big) \tilde\psi_2 :=\Big(-\partial_{xx} +\big(i\,{\Delta}\big)'  + \big(i\,{\Delta}\big)^2\Big) \tilde\psi_2 = -  k^2\, \tilde\psi_2
\end{array}\right.
\end{equation}

Let us mention some aspects of the above equations.
\begin{enumerate}
\item
If $\tilde\psi_1$ is a solution of the first equation, then its complex
conjugate  $\tilde\psi_1^* := \tilde\psi_2$ will be a solution of the second.

\item 
Each of the two equations may constitute a PT symmetric Hamiltonian
\cite{bender,Bender19,oscar15,Ho14}. For example, if the potential $\cal V$ is real and even, so that $V_i$, $i=1,2$, will also be even. The point is whether  in that case, the solutions are also PT symmetric. We can check that this is true for the example of Section 4.

\item  The equations have potentials depending on the energy of the initial Dirac-Weyl problem.
In this situation, the role of $k$ and $\varepsilon$ has been
interchanged with respect to the pure magnetic case in (\ref{pxx2}).
However, we assume that $k$ is a continuous parameter and for each value
of $k$ we should have a discrete number of energies $\varepsilon_n(k)$.

\item 
In general, we will
have complex solutions for each equation (\ref{sch}), but the spectrum given by the eigenvalues $\varepsilon$ must be real.
The Dirac-Weyl equation (\ref{px}) in the new components is obtained by applying $T^{-1}$ to (\ref{p12});
the resulting pair of equations take the form
\begin{equation}\label{pxt}
(\partial_x\sigma_1 + \Delta(x)\sigma_2)\tilde\Psi(x) = k \tilde\Psi(x), 
\quad\quad \Delta(x) = \varepsilon - {\cal V}(x)\,,\quad \tilde\Psi(x)= \left(\begin{array}{c}
\tilde\psi_1(x) 
\\[1.ex]
\tilde\psi_2(x)\end{array} \right)
\end{equation}
Therefore, the solutions must satisfy the separated equations (\ref{sch}) and also the version (\ref{pxt}) of the Dirac-Weyl equation. As we will see later, in an example, the solutions  of (\ref{sch}) can be chosen conjugate, but one must find the right complex phase coefficient so that the coupled equations (\ref{pxt}) be also satisfied.
Assume, for instance, that eq.~(\ref{sch}) is PT symmetric, according to point 2. Then,
eq.~(\ref{pxt}) is invariant under the operator $PT\sigma_3$. If both symmetries are not broken, each solution $\tilde\Psi(x)$ of both equations will also be eigenfunction of both symmetries:
\begin{equation}\label{pt}
PT\sigma_3\tilde\Psi(x)=\lambda\tilde\Psi(x),\quad
PT\tilde\psi_1(x)=\mu\tilde\psi_1(x),\quad \tilde\psi_2(x)=\tilde\psi_1(x)^*
\end{equation}
where $\lambda$, $\mu$ are the eigenvalues.
The compatibility of these conditions leads to $\mu =\pm i$. 
We will check this result later on in the example of Section 4.

\item
Remark that eq.~(\ref{pxt}) together with (\ref{sch}) have the characteristic form of SUSY-QM \cite{Cooper}, where  the intertwining operators $A^\pm$ have complex superpotentials,
$A^\pm = \partial_x \pm i \Delta(x)$. 

If we compare the  Schr\"odinger equations for pure electric fiedls (\ref{sch}) and the corresponding for pure magnetic field (\ref{pxx2}) we see that, by means of our diagonalization, we have transformed an electric problem into the form of a magnetic one including some imaginary terms. This process is the same as the ``complex Lorentz boost'' of \cite{Tan10}. The interpretation
of this kind of manipulation in terms of Lorentz transformations goes back
to \cite{lukose}. A general version of this calculation for position dependent Fermi velocity, paying attention to SUSY-QM, was also considered in \cite{Phan21}.

\end{enumerate}

We can apply this formalism to different electric potentials. We have to take into account the asymptotic conditions of the potential ${\cal V}(x)$ in order the Schr\"odinger like equations
(\ref{sch})
have square-integrable eigenfunctions corresponding to a discrete spectrum. Take, for instance the first equation of (\ref{sch}) and rewrite it as follows:
\begin{equation}
\Big(-\partial_{xx}   + (i\,{\cal V}' + 2\varepsilon {\cal V} -{\cal V}^2)\Big) \tilde\psi_1 
= -  (k^2-\varepsilon^2)\, \tilde\psi_1
\end{equation}
We consider an effective complex potential (depending on $\varepsilon$)
\begin{equation}\label{veff}
V_{\rm eff}(x) = i\,{\cal V}' + 2\varepsilon {\cal V} -{\cal V}^2
\end{equation}
and an effective energy
\begin{equation}\label{eff}
E_{\rm eff}= -(k^2-\varepsilon^2)
\end{equation}

Let us assume, for instance, that $V_{\rm eff}(x)^\pm \ \to \ 0$ when $x\ \to \ \pm\infty$. Then, a necessary condition to have an square integrable eigenfuntion is,
according to (\ref{eff}), that $|\varepsilon|<|k|$; in other words,
in the context of null mass graphene quasiparticle, the parallel momentum $k$ plays a similar role of the mass in a Dirac particle. We will see in the following example a simple case to confirm this  behaviour.

In general, the conditions to have an effective well potential are not easy to satisfy if we are interested in potentials allowing for analytic solutions.  Some examples
considered in  previous references are the following.
a) ${\cal V}(x)= \frac{e}{|x|}$\,, similar to the Coulomb potential but in one dimension \cite{portnoi4};
b) ${\cal V}(x)= \frac{e}{1+x^2}$\,, a Lorentzian type potential \cite{portnoi4};
c) ${\cal V}(x)= \tanh x$\,, a P\"oschl-Teller like potential \cite{portnoi5}.

A way to obtain partly analytic formulas of eigenfunctions is to make use of piecewise regular functions for the potentials and then apply matching conditions. This is the case of the following example that we will work out
in detail.
It will give us the main features of the discrete spectrum and a number of properties of the electric interaction: Boundary lines of the spectrum; atomic collapses produced by deepening the potential; the role of the parallel momentum $k$, or the shape and orthogonality of the complex eigenfunctions.

\subsection{Proportional electromagnetic fields: ${\cal V}\propto {\cal A}$}

In this case we will assume that both potential intensities are proportional,
\[
{\cal V}(x) = \alpha {\cal A}(x)\ \implies \ \left\{\begin{array}{ll}
W(x) = k+ {\cal A}(x)
\\[1.5ex] 
\Delta(x) =  \varepsilon-\alpha{\cal A}(x)
\end{array}\right.
\]
Then, we can write
\[
\Delta'(x) = -\alpha {\cal A'}(x),\qquad W'(x) =  {\cal A'}(x)
\] 
Thus, the  matrices $M'$ and $M^2$ will take the form  
\[
M'
= 
{\cal A'}(x)
\left(\begin{array}{cc}
1 & \alpha
\\[1.ex]
-\alpha & -1\end{array} \right):= {\cal A'}(x) M'_\alpha,
\qquad
M^2= \Big( (k+ {\cal A}(x))^2 - (\varepsilon- \alpha {\cal A}(x))^2\Big) \, I
\]
The characteristic polynomial, $\det(M_\alpha'-\lambda I)=0$, for the eigenvalues $\lambda$ of $M_\alpha'$ gives
$
\lambda^2 = 1-\alpha^2
$. 
Then, we have the following cases depending on $\alpha$:
\begin{itemize}
\item[a)]\ $ |\alpha|<1$\, \ both eigenvalues $\lambda_1\neq \lambda_2$ are real   (trigonometric  or ${\cal A}$-dominant) 
\item[b)]\ $ |\alpha|>1$ \ both eigenvalues $\lambda_1\neq \lambda_2$ are pure imaginary  (hyperbolic or ${\cal V}$-dominant)
\qquad 
\item[c)] \ $ \alpha =\pm 1$, \ both eigenvalues are zero $\lambda_1= \lambda_2=0$   (parabolic or equal ${\cal A}$-${\cal V}$)
\end{itemize}

Henceforth we will assume $0<|\alpha|<1$. The second order equation (\ref{pxx}) can be written in the form
\begin{equation}\label{em0}
-\Psi_R''(x)+\left(M'
+\Big( (k+ {\cal A}(x))^2 - (\varepsilon- \alpha {\cal A}(x))^2\Big) \, I\right) \Psi_R(x) =0
\end{equation}
Next, we diagonalize the matrix $M'(x)$ by means of the eigenvalues and eigenvectors of the constant matrix $M'_\alpha$:
\[
\lambda_{\pm} = \pm \sqrt{1-\alpha^2},
\qquad 
{\bf v}_\pm = \left(\begin{array}{c}
\alpha
\\[1.5ex] 
-1\pm \sqrt{1-\alpha^2} \end{array} \right)
\]
Sometimes we use the notation $\alpha = \sin \gamma$; in this case we have
\[
\lambda_{\pm} = \pm \cos \gamma,
\qquad 
{\bf v}_\pm = \left(\begin{array}{c}
\sin \gamma
\\[1.5ex] 
-1\pm \cos \gamma \end{array} \right)
\]
  The diagonalization is realized by means of the matrix $T$ given by
\[
T=\left(\begin{array}{cc}
\alpha & \alpha
\\[1.5ex] 
-1+ \sqrt{1-\alpha^2} & -1- \sqrt{1-\alpha^2} \end{array} \right)
\]
as follows,
\[
\Psi_R(x) = T \tilde\Psi(x),\quad 
 T^{-1} M' T={\cal A}'(x)\left(\begin{array}{cc}
 \sqrt{1-\alpha^2} & 0
\\[1.5ex] 
0 & - \sqrt{1-\alpha^2} \end{array} \right):={\cal A}'(x)D_\alpha, 
\]
Once replaced these expressions in (\ref{em0}) we get
\begin{equation}\label{em1}
-\tilde\Psi''(x)+\left({\cal A'}(x) D_\alpha +\Big(\frac{\varepsilon\alpha+k}{\sqrt{1-\alpha^2}} +
\sqrt{1-\alpha^2} {\cal A}(x))\Big)^2 -\frac{(\varepsilon +\alpha k)^2}{1-\alpha^2} \, I\right) \tilde\Psi(x) =0
\end{equation}
Let us call the upper and lower components of $\tilde\Psi(x)$ in the form $\tilde\psi^+(x)$ and $\tilde\psi^-(x)$,
respectively, then each one satisfies the equation:
\begin{equation}\label{em2}
-(\tilde\psi^{\pm})''(x)+ \left(  W^2(x)\pm W'(x) -\mu \right) \tilde\psi^{\pm}(x) =0
\end{equation}
where
\begin{equation}\label{em3}
W(x)= \frac{\varepsilon\alpha+k}{\sqrt{1-\alpha^2}} +
\sqrt{1-\alpha^2}\, {\cal A}(x),\quad
\mu = \frac{(\varepsilon +\alpha k)^2}{1-\alpha^2},
\end{equation} 
or
\begin{equation}\label{em4}
 W(x)= \frac{\varepsilon\sin \gamma+k}{\cos\gamma} +
\cos\gamma\,  {\cal A}(x),\quad
\mu = \frac{(\varepsilon +k \sin\gamma)^2}{\cos^2\gamma}
\end{equation}
All this means that the components $\tilde\psi^\pm$ are supersymmetric components of
two partner Hamiltonians \cite{kuru}.

These formulas are valid for the magnetic case where $0<|\alpha|<1$; however they can be
extended to the electric case with $|\alpha|>1$, just by including some pure
imaginary terms in (\ref{em3}):
\[
\sqrt{1-\alpha^2}= i\sqrt{\alpha^2-1}=i\,\sinh \gamma\,,\qquad \alpha= \cosh \gamma
\]
For the trigonometric case, $\alpha= \cos \gamma$, if we choose ${\cal A}(x)= \beta x$  then we obtain the effective potential
of the harmonic oscillator with a frequency depending on $\alpha$.
This kind of examples were dealt with in  \cite{lukose,roy20,Ghosh19}  and references therein; also for applications of this case to coherent states see \cite{Bautista20}.

\section{Square well potential}
Let us consider in this section the case of a pure electric square well  potential with profile defined by
\begin{equation}\label{sqw}
{\cal V}(x)= 
\left\{\begin{array}{cl}
-v_0,  \ &|x|<1 \,, 
\\[1.5ex]
 0,  \ &|x|>1
 \end{array}\right.
\end{equation}
The corresponding Schr\"odinger equation with effective potential (\ref{veff}) is:
\begin{equation}\label{e1s}
-\tilde\psi_1''(x) + 
V_{\rm eff}(x)
=-(k^2 - \varepsilon^2)\tilde\psi_1(x)
\end{equation}
Here, the complex effective potential has the following form:
\begin{equation}\label{epot1}
V_{\rm eff}(x) = 
\left\{ \begin{array}{l} 
-i v_0 \delta(x+1)+iv_0 \delta(x-1)-2 v_0  \varepsilon-v_0^2, \quad |x|\leq 1,
\\[1.5ex]
  0,\quad |x|>1
\end{array}\right.
\end{equation}
For this potential we have three different regions: $x<-1$,  $-1<x<1$,   and $x>1$. In the first and third region the potential  vanishes while, in the second one is a constant $(-2 v_0  \varepsilon-v_0^2)$. Each region is separated from its neighbour by a  Dirac's delta with imaginary intensity. Next, we write the solutions corresponding to bound states within each of these regions:
\begin {itemize}
\item $x<-1$:
\begin{equation}\label{e1s1}
\begin{array}{ll}
-\tilde\psi_{1,I}''(x) =-p^2 \tilde\psi_{1,I}(x),\qquad & p^2=k^2-\varepsilon^2>0
\\[1.5ex]
\tilde\psi_{1,I}(x) =A e^{px},\qquad &p=\sqrt{k^2-\varepsilon^2}>0
\end{array}
\end{equation}
\item $-1<x<1$:
\begin{equation}\label{e1s2}
\begin{array}{ll}
-\tilde\psi_{1,II}''(x) +(-2 v_0  \varepsilon-v_0^2)\tilde\psi_{1,II}(x)=-p^2 \tilde\psi_{1,II}(x),\qquad & q^2=2 v_0  \varepsilon+v_0^2-p^2>0
\\[1.5ex]
\tilde\psi_{1,II}(x) =C e^{iqx}+D e^{-iqx},\qquad &q=\sqrt{(\varepsilon+v_0)^2-k^2}>0
\end{array}
\end{equation}
\item $x>1$:
\begin{equation}\label{e1s3}
\begin{array}{ll}
-\tilde\psi_{1,III}''(x) =-p^2 \tilde\psi_{1,III}(x),\qquad &p^2=k^2-\varepsilon^2>0
\\[1.5ex]
\tilde\psi_{1,III}(x) =F e^{-px},\qquad &p=\sqrt{k^2-\varepsilon^2}>0
\end{array}
\end{equation}
\end{itemize}

\begin{figure}[h!]\label{energy1}
\begin{center}
\includegraphics[scale=0.4]{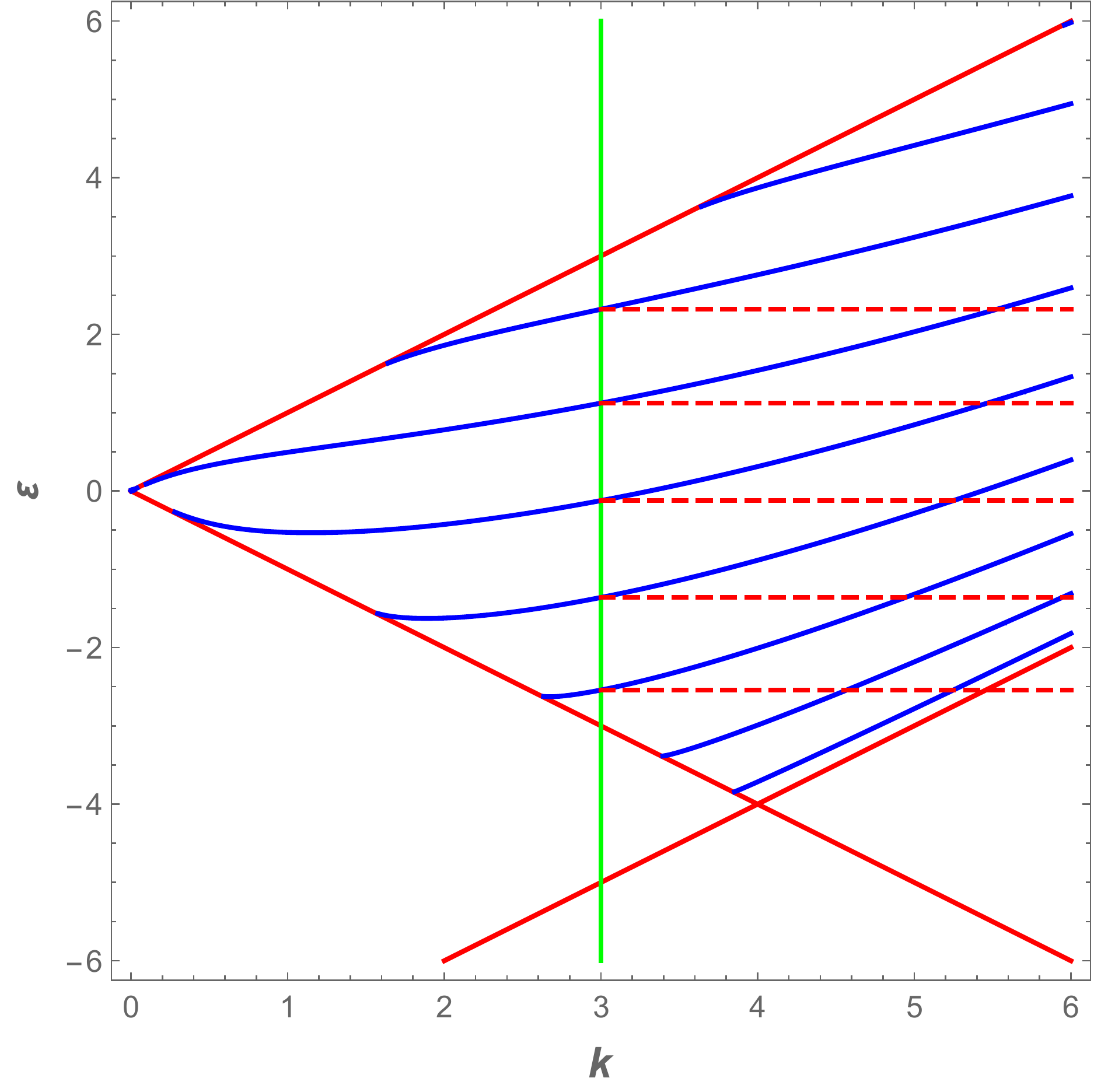}
\hskip0.4cm
\includegraphics[scale=0.4]{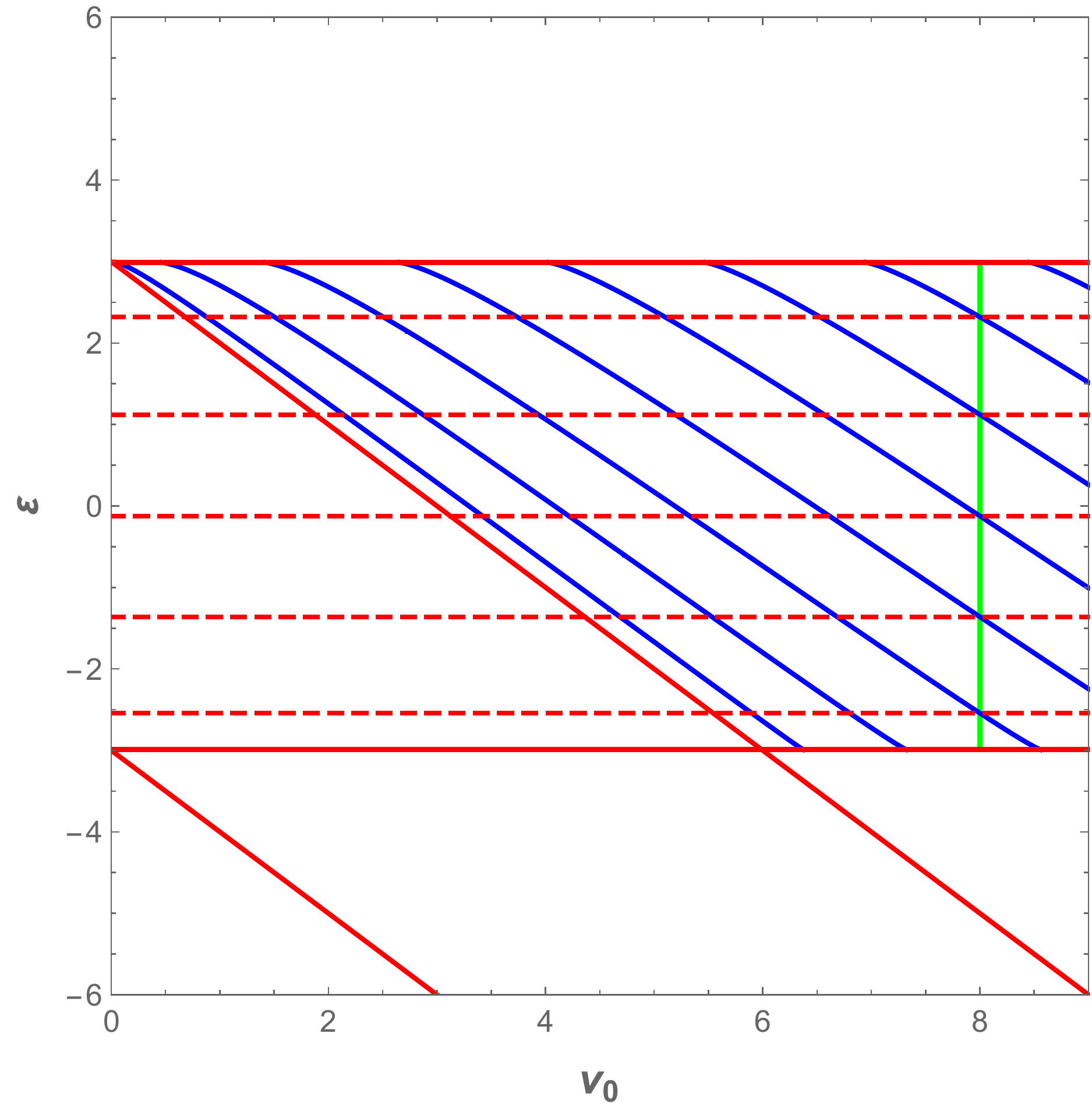}
\caption{(left) Graphic 
of  energy eigenvalues as a function of $k$ for a fixed value $v_0=8$ and the border lines $\varepsilon=\pm k$ and $\varepsilon=-8+ k$ (the green vertical line is for $k=3$.)
 (right)  Graphic of energy $\varepsilon$ versus the depth $v_0$ of the well for $k=3$. The green line corresponds to $v_0=8$; the spectrum of the two vertical green lines should coincide (as shown by the dashing red lines). \label{f2}}
\end{center}
\end{figure}

These solutions  satisfy continuity conditions at $x=-1$ and $x=1$.
\begin{equation}\label{cn1}
\begin{array}{ll}
\tilde\psi_{1,I}(-1) =\tilde\psi_{1,II}(-1),\qquad &\tilde\psi_{1,I}'(-1)=\tilde\psi_{1,II}'(-1)+i v_0\tilde\psi_{1,I}(-1)
\\[1.5ex]
\tilde\psi_{1,II}(1) =\tilde\psi_{1,III}(1),\qquad &\tilde\psi_{1,II}'(1)=\tilde\psi_{1,III}'(1)-i v_0\tilde\psi_{1,II}(1)
\end{array}
\end{equation}
From these conditions we have four equations for $A, C, D, F$:
\begin{equation}\label{s1}
\begin{array}{c}
A e^{-p}-C e^{-iq}-De^{iq}=0
\\[1.5ex]
A (pe^{-p}-iv_0e^{-p})-C (iq)e^{-iq}+D(iq)e^{iq}=0
\\[1.5ex]
C e^{iq}+De^{-iq}-Fe^{-p}=0
\\[1.5ex]
C (iq)e^{iq}+D(-iq)e^{-iq}+F (pe^{-p}+iv_0e^{-p})=0
\end{array}
\end{equation}
This is a  homogeneous system  for the unknown parameters $A, C, D, F$. The solutions exist when the determinant of the matrix ${\cal M}$ of the coefficients in (\ref{s1}) vanishes, $\det({\cal M})=0$. Then, we get the relation between $\varepsilon$, $k$ and $v_0$ from this secular equation:

\begin{equation}\label{det}
\sqrt{k^2-\varepsilon^2} \sqrt{(\varepsilon+v_0)^2-k^2} \cos \left(2 \sqrt{(\varepsilon+v_0)^2-k^2}\right)-\left(\varepsilon (\varepsilon+v_0)-k^2\right) \sin \left(2 \sqrt{(\varepsilon+v_0)^2-k^2}\right)=0
\end{equation}
The characteristics of energy  values, $\varepsilon$, as well as eigenfunctions of the solutions of this complex Schr\"odinger equation will be discussed along the following figures.

Fig.~\ref{f2} (left) shows the graphic of $\det({\cal M})=0$ for a range of $k$ and $\varepsilon$ once fixed $v_0=8$. The curves in blue represent the values of the energy.  The Red lines $\varepsilon=\pm k$, see (\ref{e1s1}), and $\varepsilon=\pm k-v_0$,  according to (\ref{e1s2}),  determine the domains where the solutions may exist.  In particular, the solutions on $\varepsilon=\pm k$  correspond to the critical values for the bound state energies. 
 In our problem, only when $\varepsilon + v_0 > k$ and $\varepsilon < k$ are satisfied, there will be solutions of the previous equations. This graphic can be extended to the negative values $k<0$ in a symmetric way.

It is quite significant the plot of the $\varepsilon$ dependence on the potential $v_0$ for fixed values of $k$ where the colapses of negative bound states in the negative sea can be appreciated
\cite{katnelson,moldovan}. This is shown in the plot to
the right of Fig.~\ref{f2}. We pay attention on the five points of the spectrum corresponding to $v_0=8$ (represented by a green line; the points of the spectrum are given by its five intersections with the blue curves; they coincide with the spectrum of the `equivalent' green line --same $v_0$, same $k$--, on the left  of Fig.~\ref{f2}, as it is shown by the horizontal dashing lines). In this case the lowest
state of the spectrum is not a fundamental state due to two previous collapses
(as shown in Fig.~\ref{f2}, right), so that in fact, this is the second excited state. The red lines are boundaries of solutions, as mentioned earlier.

\begin{figure}[h!]
\begin{center}
\includegraphics[scale=0.4]{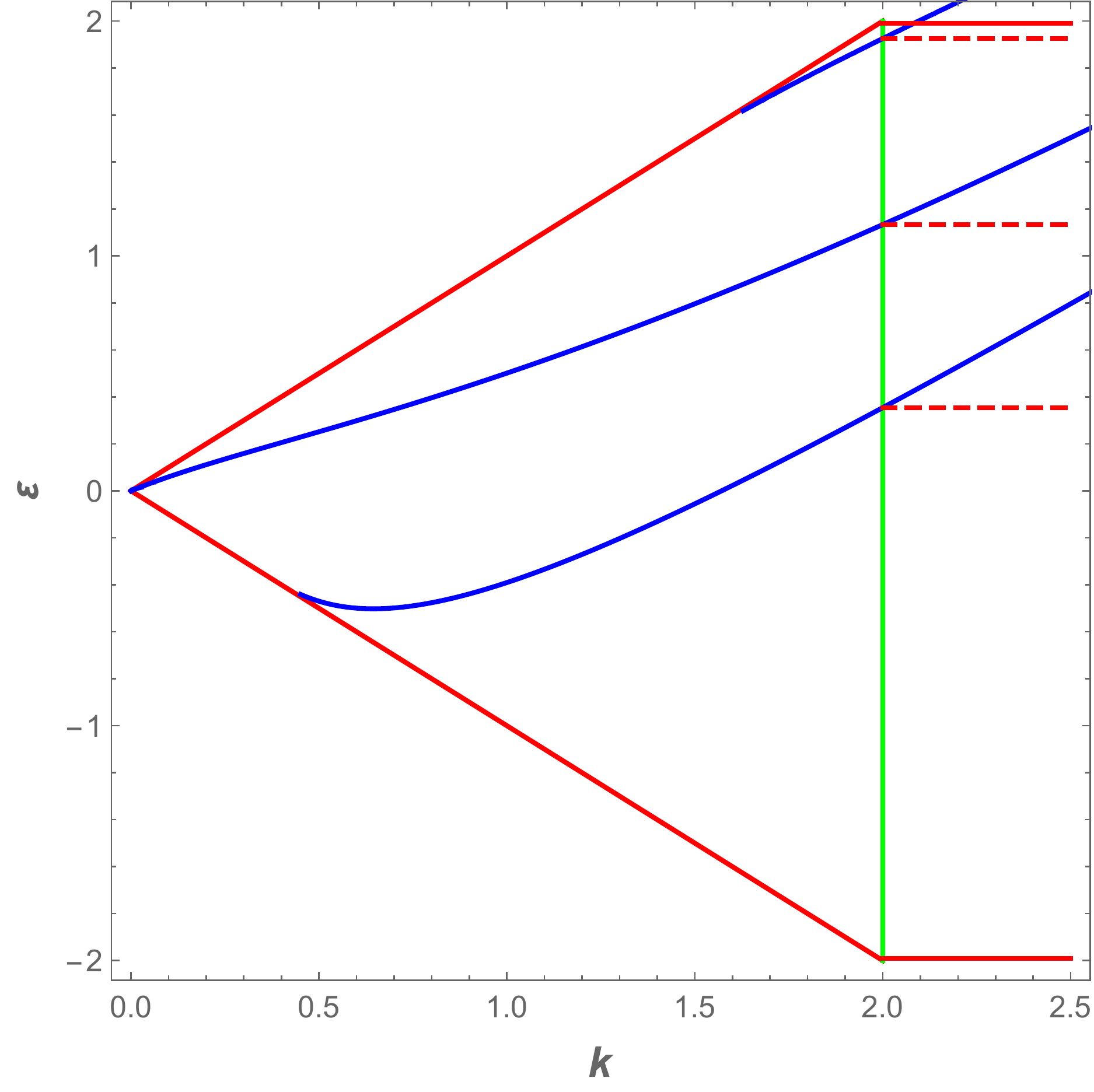}
\hskip0.4cm
\includegraphics[scale=0.4]{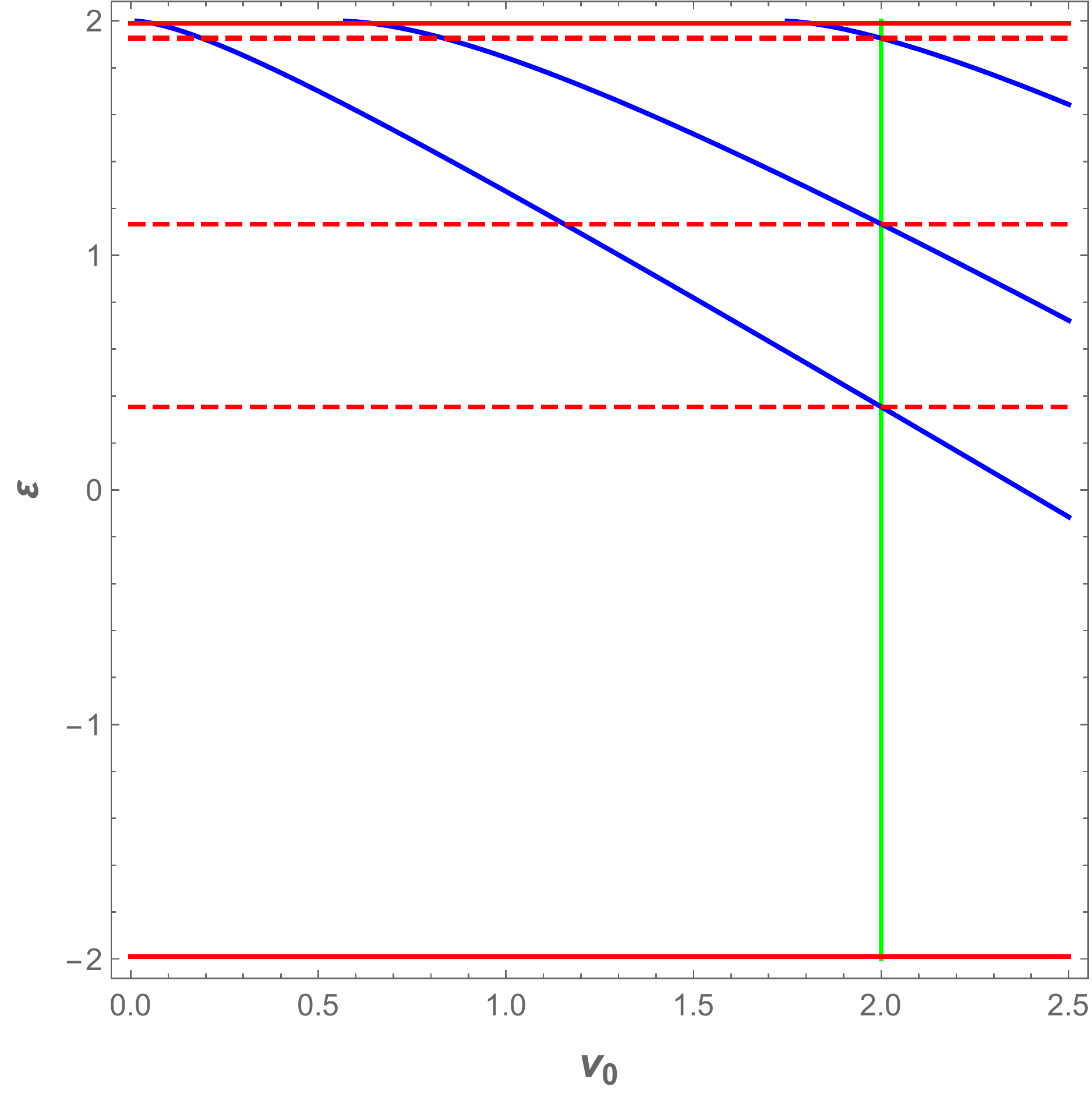}
\caption{(left) Graphic of ${\rm Im}(\det{\cal M})=0$  (energy eigenvalues) for a fixed value $v_0=2$ and the border lines $\varepsilon=\pm k$;
and (right)  the graphic of energy $\varepsilon$ versus the depth $v_0$ of the well  for $k=2$.\label{f3}}
\end{center}
\end{figure}

\begin{figure}[h!]
\begin{center}
\includegraphics[scale=0.35]{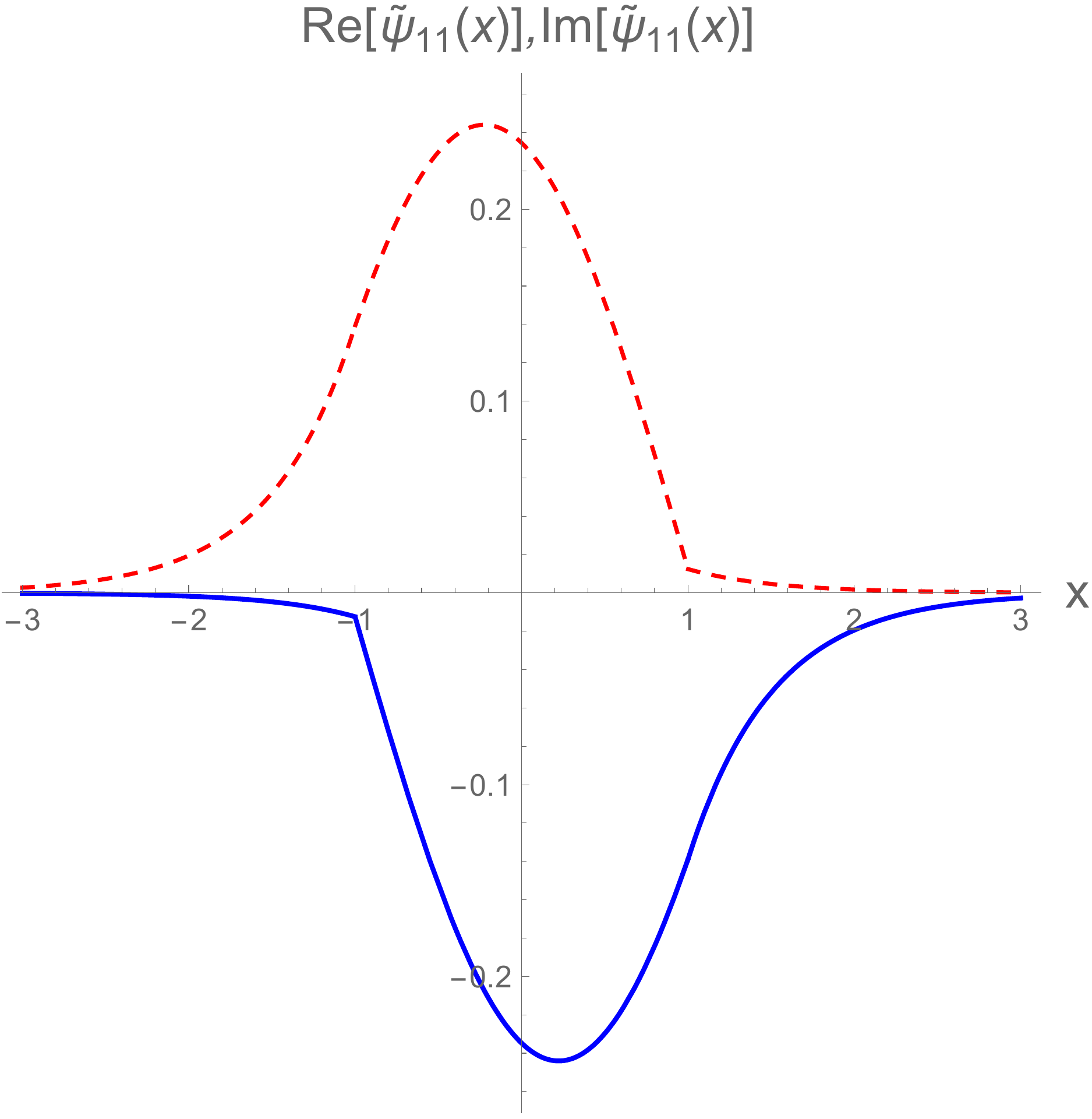}
\includegraphics[scale=0.35]{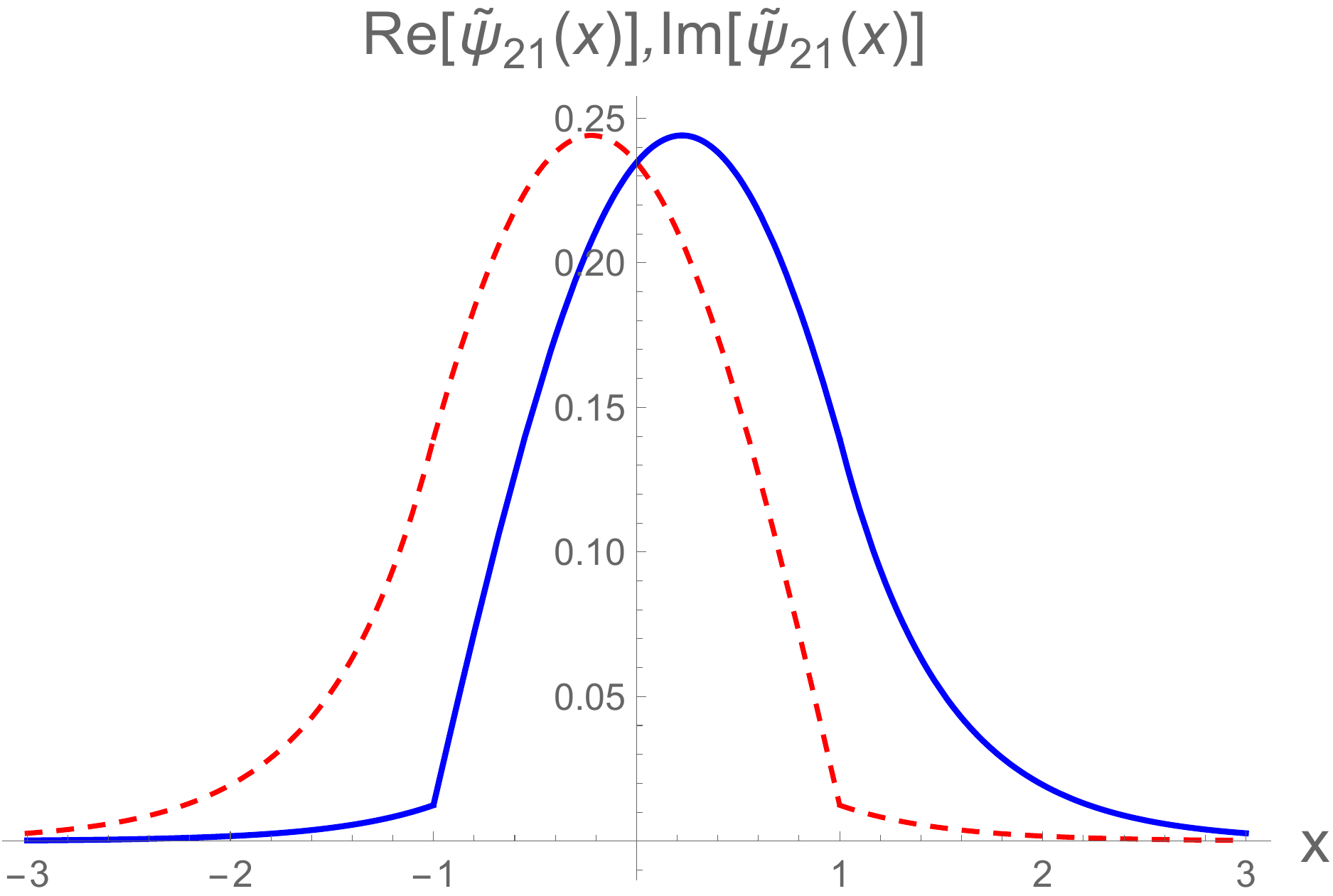}
\caption{The graphic of real part (dotted) and  imaginary part (continuous) of the ground spinor $(\tilde\psi_{11}(x),\tilde\psi_{21}(x))$  for $v_0=2$, $k=2$ and $\varepsilon_1=0.354274$.\label{f4}}
\end{center}
\end{figure}

\begin{figure}[h!]
\begin{center}
\includegraphics[scale=0.4]{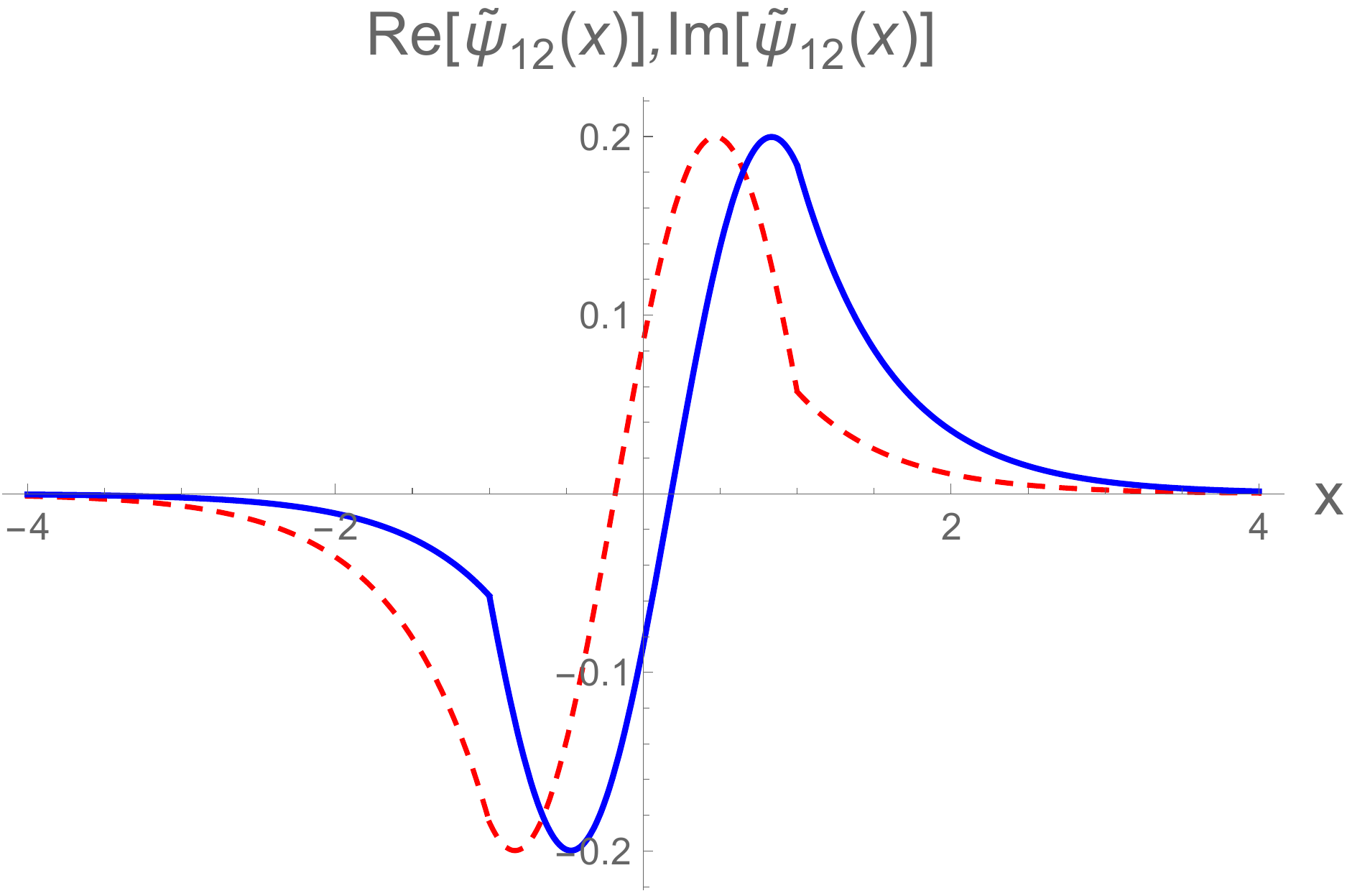}
\includegraphics[scale=0.4]{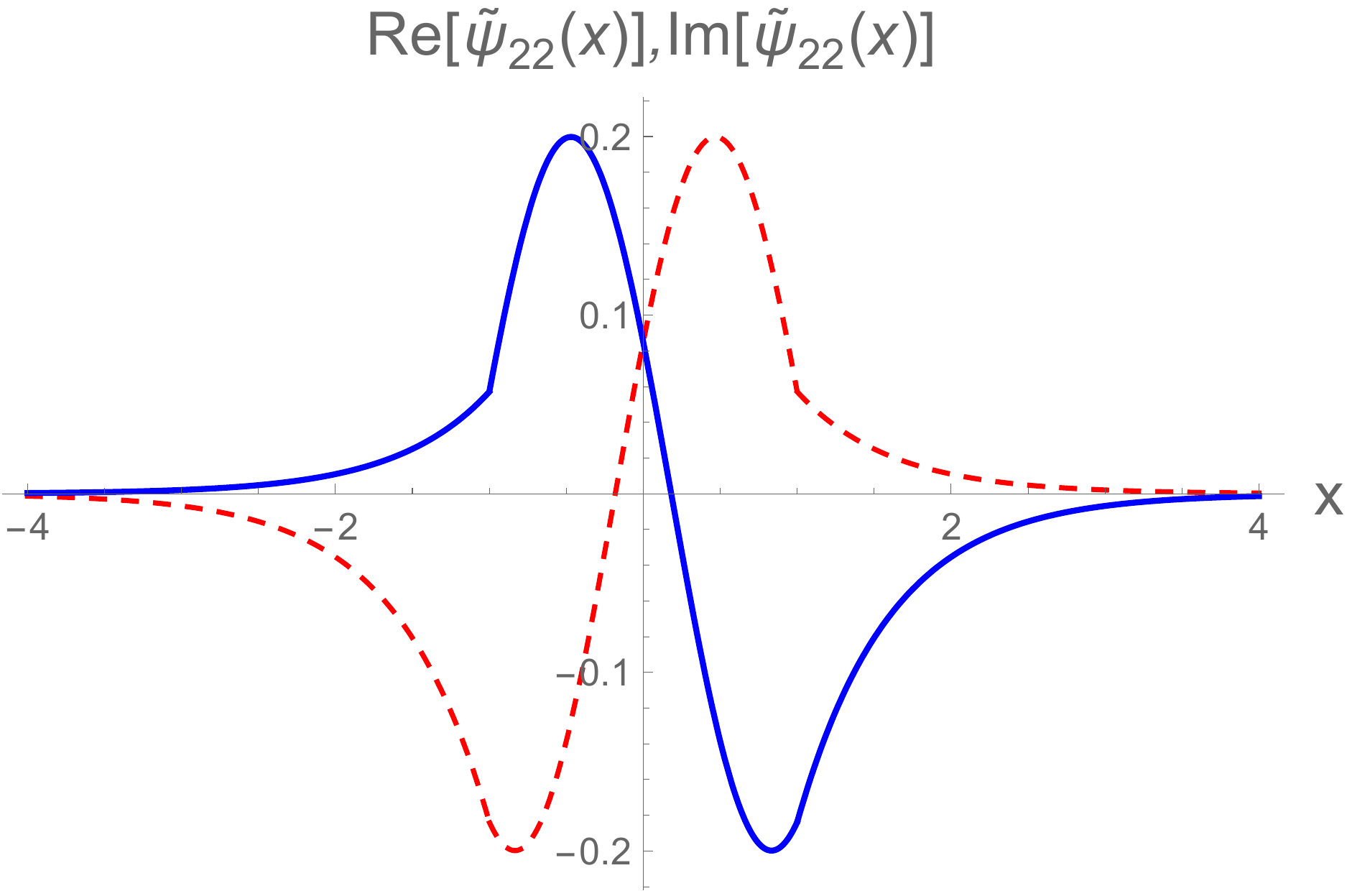}
\caption{The graphic of real part (dotted) and  imaginary part (continuous) of the second solution $(\tilde\psi_{12}(x),\tilde\psi_{22}(x))$  for $v_0=2$, $k=2$ and $\varepsilon_2=1.13356  $.\label{f5}}
\end{center}
\end{figure}

\begin{figure}[h!]
\begin{center}
\includegraphics[scale=0.4]{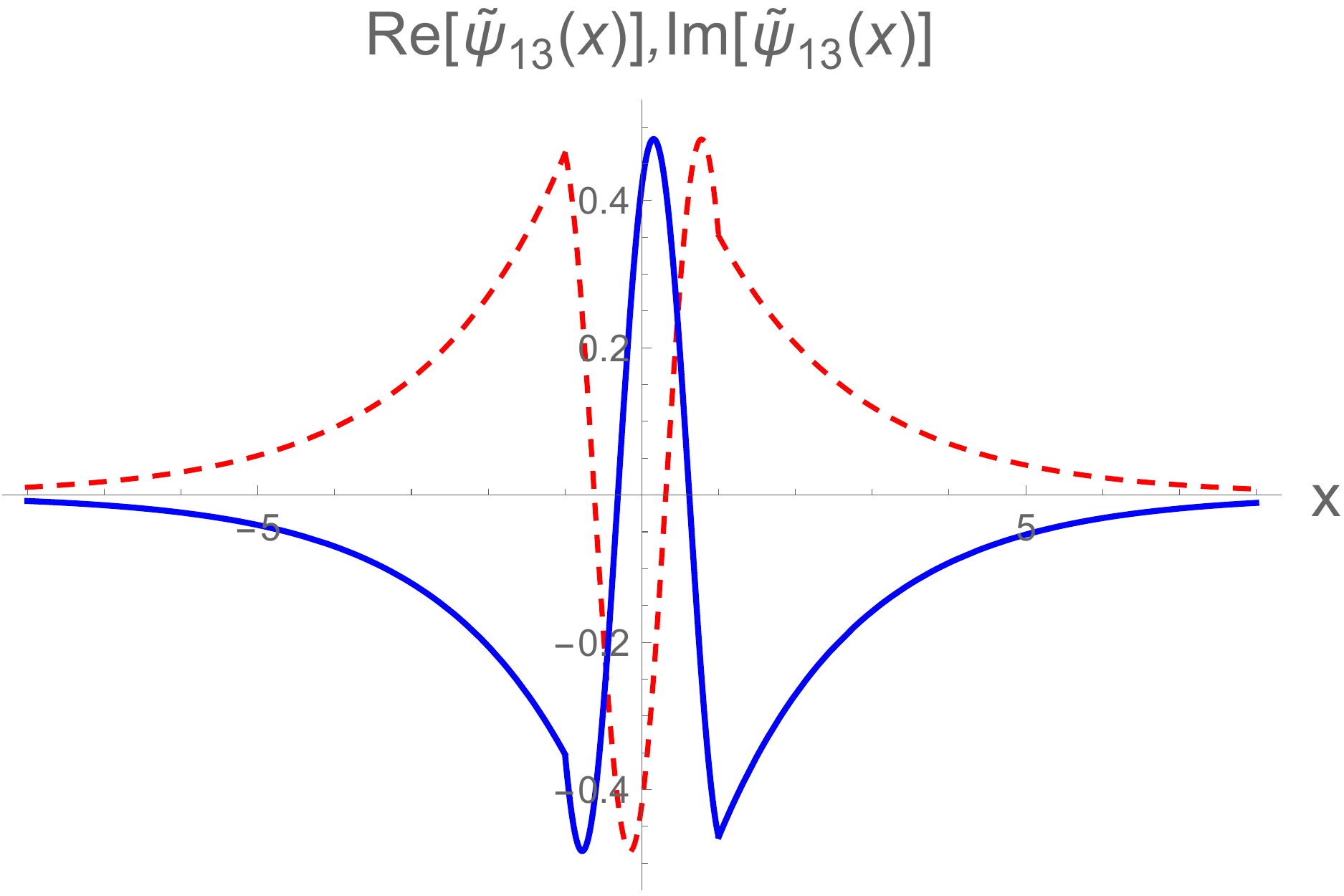}
\includegraphics[scale=0.4]{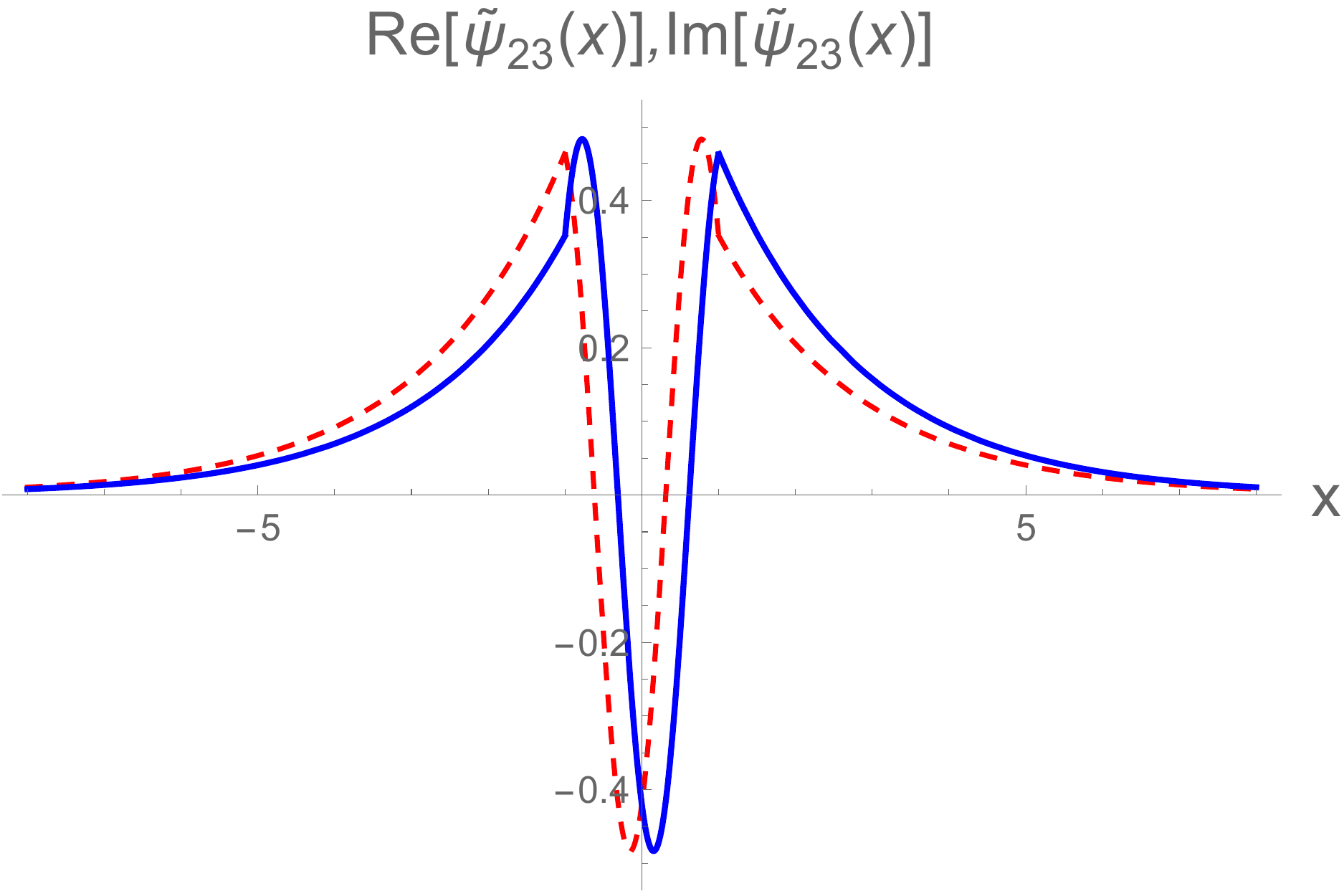}
\caption{The graphic of real part (dotted) and  imaginary part (continuous) of the third level spinor $(\tilde\psi_{13}(x),\tilde\psi_{23}(x))$ for $v_0=2$, $k=2$ and $\varepsilon_3=1.92583 $.\label{f6}}
\end{center}
\end{figure}

Fig.~\ref{f3} is similar to Fig.~\ref{f2}, but in this case $v_0=2$ for the left graphic, while $k=2$ for the right one. 
The states of the three points of the spectrum determined by the green lines have been
obtained and plotted in Figs.~\ref{f4}-\ref{f6}. These complex eigenfunctions have some basic properties: i) For each spinor the two components are complex conjugate of each other: $\tilde\psi_1(x) = \tilde\psi_2(x)^*$. ii) The solutions $\tilde\Psi$ 
corresponding to different energy levels are orthogonal and therefore also the spinors $\tilde \Psi$ are orthogonal. The orthogonality is defined as follows. Let $\tilde \Psi^a=(\tilde \psi^a_1,\tilde \psi^a_2)$ the spinor components of an state $\tilde \Psi^a$ with energy $E^a$,
and $\tilde \Psi^b$ that one with energy $E^b\neq E^a$. Then, the orthogonality is obtained from the standard formula by replacing the functions according to (\ref{p12}),
\[
\langle \tilde \Psi^a, \tilde \Psi^b\rangle := \int_{-\infty}^{\infty}(\tilde \psi^a_2 \tilde \psi^b_1+\tilde \psi^a_1\tilde \psi^b_2) d x=\frac{1}{2}\int_{-\infty}^{\infty} (\psi^a_1  \psi^b_1+ \psi^a_2 \psi^b_2) d x=0
\]
iii) Remark that $\tilde\psi_1$ and $\tilde \psi_2$ besides  satisfying  (\ref{sch}) and being complex conjugate of each other  they have also to  satisfy  equations  (\ref{pxt}).
So, it may be necessary to multiply $\tilde\psi_1$ and $\tilde\psi_2$ by conjugate phases.  From the graphics of Figs.~\ref{f4}-\ref{f6} it can be seen that 
these solutions implement $PT$ symmetry.
In other words, the energy eigenfunctions $\tilde\psi_k(x)$ are also eigenfunctions
of the $P\,T $ operator (made of complex conjugation $T$ and $x$ parity $P$) with
eigenvalues $\pm i$, as it should be, after (\ref{pt}):
\[
P\,T \tilde \psi_k(x) := \tilde\psi_k(-x)^* = (\pm i) \tilde\psi_k(x),\qquad  k=1,2
\]
We should point out that  Figs.~\ref{f4}-\ref{f6} correspond to ground, first and second excited states, respectively. Indeed, they have zero, one and two nodes corresponding to such kind of states. This implies that there have been no collapses before $v_0=2$, as seen in Fig.~\ref{f3} (right).

Using $\tilde\psi_{1,2}$, the solutions of Dirac Weyl equation given by (\ref{psiy}) can be obtained. In order  to do this, first we use equation (\ref{p12}) to get $\psi_{1,2}$ in terms of $\tilde\psi_{1,2}$.  Thus, the spinor solutions of Dirac-Weyl equation are:

\begin{equation}\label{psiyn}
\Psi_n(x,y)=2 e^{i k \,y}
\left(\begin{array}{c}
{\rm Re} \tilde\psi_{1n}(x)
\\[1.ex]
-i\,{\rm Im} \tilde\psi_{1n}(x) 
\end{array}\right)
\end{equation}

\begin{figure}[h!]
\begin{center}
\includegraphics[scale=0.35]{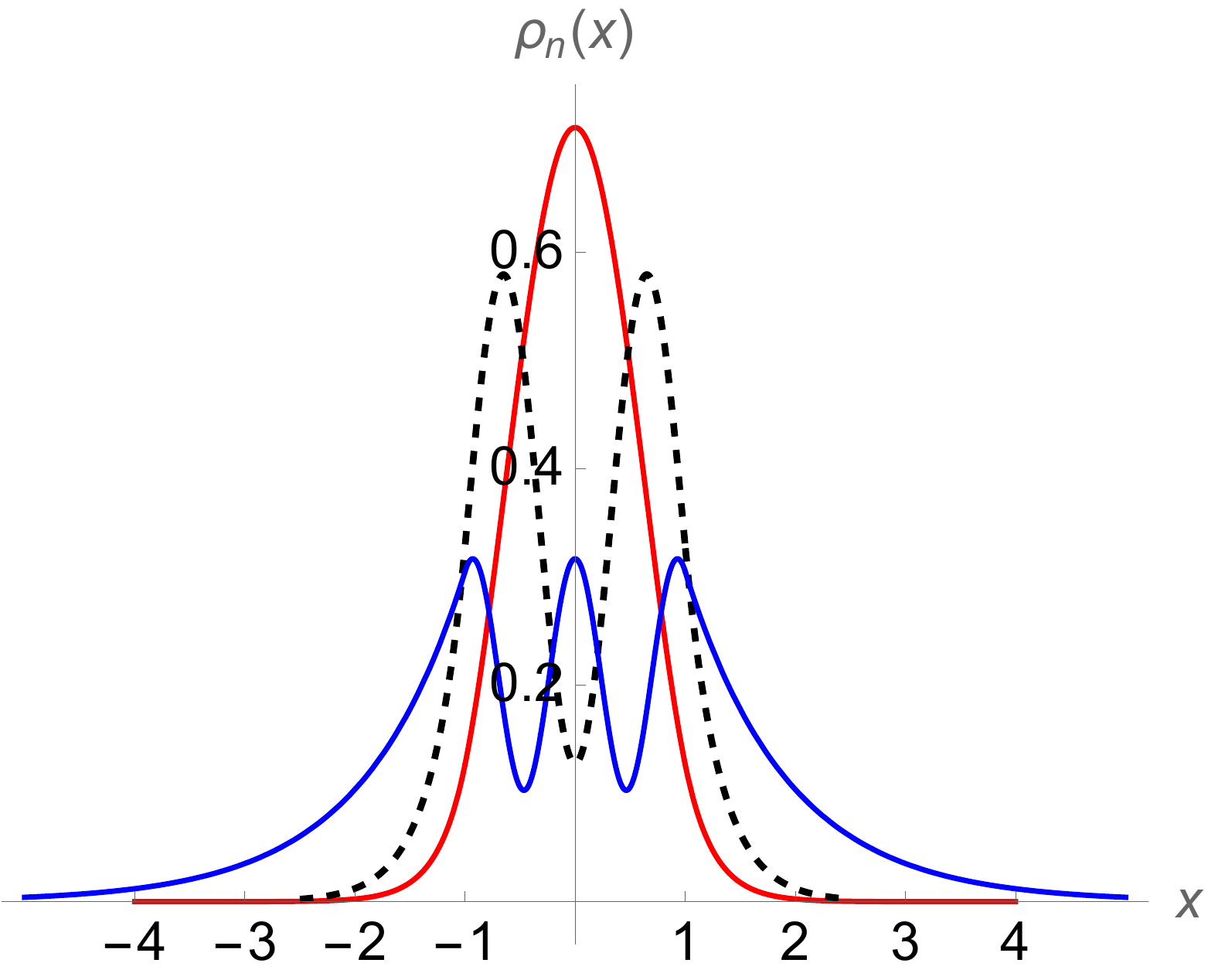}
\includegraphics[scale=0.4]{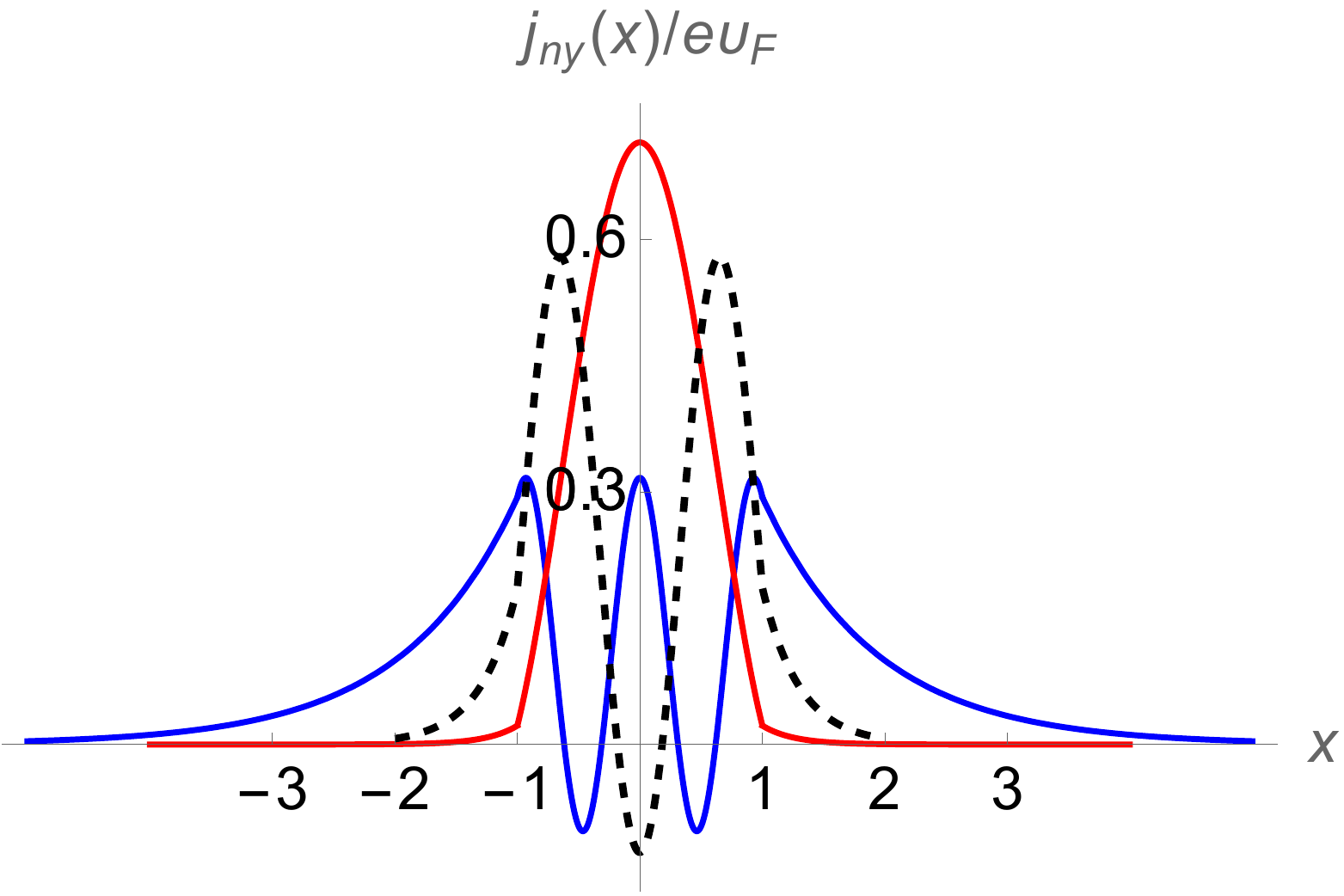}
\caption{Graphics of probability (left) and current (right) densities
for the three bound states of Figs.~\ref{f4}, \ref{f5} and \ref{f6} of the system with $v_0=2$, $k=2$ .
The ground state corresponds to the red curve, the first excited state is in black dotted, and the second excited state is in blue.
\label{f7}
}
\end{center}
\end{figure}

Then, probability density 
\begin{equation}\label{ro}
\rho_n=|\Psi_n(x,y)|^2=4 (({\rm Re} \tilde\psi_{1n}(x))^2+({\rm Im} \tilde\psi_{1n}(x))^2)
\end{equation}
and current density
\begin{equation}\label{jxy}
J_{nx}=\Psi_n^{\dagger}(x,y)\sigma_x\Psi_n(x,y)= 0,
\qquad J_{ny}=\Psi_n^{\dagger}(x,y)\sigma_y\Psi_n(x,y)= -8({\rm Re} \tilde\psi_{1n}(x))({\rm Im} \tilde\psi_{1n}(x))
\end{equation}
are found. The graphics of them for fixed values of $k$ and $v_0$ can be seen in 
Fig.~\ref{f7}.

\section{Conclusions}

In this paper we have investigated the conditions to obtain analytic solutions of
Dirac-Weyl equation in graphene under electric and magnetic fields. We have obtained
a classification in three cases, which was implicit in many references, but we
hope that our approach includes many of them in a simple and general way. 

An example,
for a pure electric potential, has been examined in detail, where we found some 
interesting properties such as: i) the relation of this problem with complex effective Schr\"odinger 
potentials; ii) the components of the spinor satisfy the formalism of supersymmetric quantum mechanics but with complex superpotentials; iii) the spectrum and bound states have been computed;  
the effective Hamiltonian is PT symmetric and we show that the eigenfunctions implement this
symmetry. Therefore, the PT symmetry is not broken which leads to a real spectrum. The eigenfunctions constitute an orthogonal system.

All these properties are quite appealing, specially the connection of electric potentials with complex effective Schr\"odinger equations. We have also shown different plots of energy-momentum $k$ of the particle in the $y$ direction, as well as energy-depth of the potential well. These two kinds of energy plots are complementary, but they are seldom present in the literature.

We also plotted the form of some eigenfunctions together with the probability and current densities which we hope will enlighten some points of the presentation, in particular the symmetries (spacial and PT).

\section*{Acknowledgments}

We appreciate the support of the QCAYLE project, funded by the European Union--NextGenerationEU, and PID2020-113406GB-I0 project funded by the MCIN of Spain.
\c{S}.~K. thanks Ankara University and the warm hospitality of the Department of Theoretical Physics of the University of Valladolid, where part of this work has been carried out, and to the support of its GIR of Mathematical Physics.

\end{document}